\documentclass[11pt]{iopart}

\usepackage{graphicx}
\usepackage{iopams}

\def\a{\alpha}
\def\b{\beta}

\def\d{\delta}
\def\l{\lambda}

\def\ket{\rangle}
\def\bra{\langle}
\def\r{\rho}
\def\th{\theta}

\def\z{\zeta}
\def\round{\partial}

\def\ri{\rightarrow}

\begin{document}

\title [The effect of detachment and attachment to a kink motion in ASEP]
 {The effect of detachment and attachment to a kink motion in the
 asymmetric simple exclusion process}
\author{Tetsuya Mitsudo\footnote[1]{mitsudo@yukawa.kyoto-u.ac.jp} and Hisao Hayakawa\footnote[2]{hisao@yukawa.kyoto-u.ac.jp}}

\address{Department of Physics, Yoshida-South campus, Kyoto University, Sakyo-ku, Kyoto, Japan, 606-8501}

\begin{abstract}
We study the dynamics of a kink in a one-lane asymmetric simple
 exclusion process with detachment and attachment of the particle at arbitrary sites.
For a system with one site of detachment and attachment 
we find that the kink is trapped by the site, 
and the probability distribution of the kink position is described by
 the overdumped Fokker-Planck equation with a V-shaped potential.
Our results can be applied to the motion of a kink in arbitrary number of
 sites where detachment and attachment take place.
When detachment and attachment take place at every site,
 we confirm that the kink motion obeys the diffusion in a harmonic
 potential.
We compare our results with the Monte Carlo simulation, and check the
 quantitative validity of our theoretical prediction of the diffusion
 constant and the potential form.
\end{abstract}

\section{Introduction}

An asymmetric simple exclusion process (ASEP) is a simple
nonequilibrium stochastic lattice model of transport process.
ASEP is defined on a one-dimensional lattice, and particles hop
to the neighbouring site when the site is empty.
The transportation of particles is induced by the asymmetric hopping
of the particles.
Recently much attention is paid to ASEP \cite{GS1,SZ,SP}, not only
because there exists the
exact solution under the open boundary condition \cite{DEHP,GS3,SS,USW},
but also it is applicable to various transportation phenomena. 
In particular, the uni-directional ASEP which is called totally asymmetric
simple exclusion process (TASEP) has been
studied extensively, because (i) TASEP is the simplest ASEP and (ii)
TASEP keeps the essence of nonequilibrium transport processes such as
the exclusion interaction between particles and the drift of particles.
In fact, TASEP may be regarded as  a simplified model of traffic flow
\cite{HELB}.

We can also apply ASEP to biological problems.
TASEP is first introduced as a model to explain the process of creation
of the messenger RNA \cite{MGP}.
The model with detachment and attachment on each site in
TASEP with open boundary conditions is
known as PFF model \cite{PFF}, named after the authors' 
names: Parmeggiani, Franosch and Frey.
An extended model of the PFF model is used to describe an intra-cellular
transport of the single-headed kinesin (KIF1A) motor \cite{NOSC}.

In the open boundary ASEP, we can draw the phase diagram by the
incoming rate and outgoing rate at the boundaries.
On the boundary 
between the low density phase and high density phase, 
it is known that the kink between a sparse region and a jammed 
region obeys Brownian motion. 
This kink motion in one-lane ASEP is also studied in terms of the
domain wall theory \cite{KSKS,SA,TMH} and the second class particle
\cite{FKS,FE,FF}.
The kink motion in a two-lane TASEP is studied in
ref.\cite{TH}, and
the kink motion in the PFF model is studied in ref.\cite{EJS,JS,MB,PRWKS}. 
It is notable that the kink is trapped by a harmonic potential in the
PFF model \cite{EJS,JS}.

In this paper, we discuss the kink motion of TASEP with detachment
and attachment.
Our method can be used for any number of sites of detachment and the
attachment.
The effect of detachment of particles at the middle site of the system was
studied in ref.\cite{MK} by dividing the system into two systems of TASEP,
though the kink motion was not discussed there.
We demonstrate that one site of detachment and attachment attracts the 
kink where attractive potential is a linear function of
distance from the site of detachment and attachment.
Even when we generalize the model with many sites of detachment and
attachment, the kink feels the linear combination of the linear potentials
for one-site of detachment and attachment.
Though the analysis of linear combination of linear potential we derive
the harmonic potential in PFF model.

The organization of this paper is as follows.
We first introduce the model with detachment and attachment at one
site and briefly review previous studies on the kink motion in TASEP
without detachment and attachment in section 2.
We present a theory of a kink motion in TASEP with detachment and
attachment at one site, and extend our analysis to the kink motion in
TASEP with detachment and attachment at many sites including the PFF
model.
In section 3, we compare our analysis with the Monte-Carlo
simulation and confirm the quantitative agreement between the theory and
the simulation.
Finally, we give concluding remarks.

\section{The kink motion}

\subsection{A model for one site of detachment and attachment}

Let us explain TASEP model with detachment and attachment of particles
at one site under the open boundary conditions. 
TASEP is defined on a one-dimensional lattice of $L$ sites.
Each particle hops forward when the site in front is empty. 
Here we choose right as the drift direction of particles.
The open boundary condition is specified by a particle attachment rate
$\a$ at
the left end of the system and a particle detachment rate $\b$ at
the right end.
Here, the particle is also detached by the
rate $w_d$ and attached by the rate $w_a$ at the site $x_0$. 
TASEP with detachment at one site has first been introduced in the
ref.\cite{MK}.
They divide the system into two systems of TASEP by the site of the
detachment, and introduce the effective hopping rate to connect the two
systems at the site $x_0$ which is fixed to the middle of the system to
include the effect of detachment.
However, we can not explain the kink localization near $x_0$ by introducing the effective
hopping rate between two systems because the kink moves under the
reflective boundary condition of a virtual boundary in one subsystem
where the kink cannot cross through the virtual boundary.
Thus we need another method to explain the effect of
detachment and attachment to the kink motion.

The Brownian motion of the kink is characterized by two parameters, the
drift velocity $V_{T}$ and the diffusion constant $D_{T}$.
In the case of TASEP, $V_{T}$ and $D_{T}$ are written as \cite{FF}
\begin{equation}
\label{vandd}
 V_{T}=1-\l_{\ell}-\l_r \quad , \quad D_{T}=\frac{\l_{\ell}(1-\l_{\ell})+\l_r(1-\l_r)}{2(\l_r-\l_{\ell})},
\end{equation}
where $\l_{\ell}$ is the density in the left of the kink and $\l_r$ is
the density in the right of the kink which satisfy $\l_{\ell}<\l_r$.
Thus once $\l_{\ell}$ and $\l_r$ are known, the motion of the kink can be
determined.

It is known that the Brownian motion in a potential $U(x)$ is written by
the Langevin equation
\begin{equation}
\label{langevin}
\frac{\rmd x}{\rmd t}=-\frac{\partial
  U(x)}{\partial x}+\z(t),
\end{equation}
where $\z(t)$ is the random force satisfies
\begin{equation}
 \bra \z(t) \ket=0 \qquad \bra \z(t)\z(0) \ket=2D\d(t),
\end{equation}
where $\d(t)$ is Dirac's delta function.
The corresponding Fokker-Plank equation to eq.(\ref{langevin}) is  
\begin{equation}
\label{fokk}
 \frac{\round P(x,t)}{\round t}=\frac{\round}{\round x}\left(\frac{\round
						   U(x)}{\round
						   x}+D\frac{\round}{\round
						   x}\right)P(x,t),
\end{equation}
where $P(x,t)$ is the probability distribution of the Brownian particle
at position $x$ at time $t$.
Equation (\ref{fokk}) has the steady solution $P_{st}(x)$
\begin{equation}
\label{eqsolb}
 P_{st}(x)=C \exp\left[-\frac{U(x)}{D}\right],
\end{equation}
where $C$ is the normalization constant.
We expect that this well-known result can be used to describe the steady
distribution of the kink in our model.

For TASEP without detachment and attachment, $\l_{\ell}$ and $\l_r$ satisfy
\begin{equation}
\label{fden}
 \l_{\ell}=\a \quad , \quad \l_r=1-\a
\end{equation}
for $\a=\b<1/2$.
From eq.(\ref{vandd}), $V_{T}$ and $D_{T}$ are respectively given by 
\begin{equation}
\label{difft}
 V_{T}=0 \quad , \quad D_{T}=\frac{\a(1-\a)}{1-2\a}.
\end{equation}

If there is a site with detachment and attachment, the phase diagram
may be modified from that of the original TASEP.
However the difference is expected to be small for small rate of the
detachment and attachment ($w_a\sim w_d \ll 1$).
Thus the kink picture may be still valid for $\a=\b\ll 1/2$.

Let us consider the kink motion in TASEP with detachment and attachment
at one site.
If there is no detachment and attachment of particles, $\l_{\ell}$ and $\l_r$ 
are given by eq.(\ref{fden}).
However, as a result of detachment and attachment the density is
deviated from eq.(\ref{fden}).
Let us consider the deviation of the density for two cases :(a) the site of the
detachment and attachment is in the
low density region of the kink and when (b) the site of detachment and
attachment is in the high density region of the kink.

\begin{figure}
\begin{center}
\includegraphics[scale=0.5]{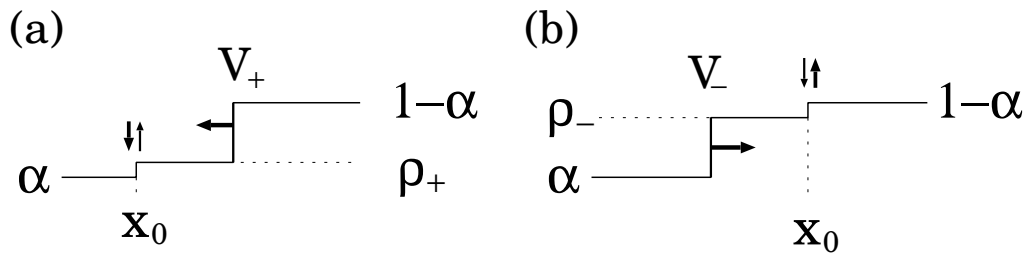}
\caption{A schematic picture of the kink motion in TASEP with one site
 of detachment and attachment.The figure (a) shows the case when $x_0$
 is in the low density region of the kink, and the figure (b) shows the
 case when $x_0$ is in the high density region of the kink.
The line shows the density profile.}
\label{fig1}
\end{center}
\end{figure}

Let us think of the case (a).
The density in the left of the kink position $x_0$ is not affected by
detachment and attachment at $x_0$ but the density in the right of
$x_0$ becomes higher as in the Fig.\ref{fig1}(a).
This situation takes place because attachment process
is likely for low-density region in the one-way motion of the particles.
We introduce $\r_+$ as the density in the right of $x_0$ when $x_0$ is in
the low density region of the kink.
Thus $\l_{\ell}$ should be $\l_{\ell}=\r_+$ while $\l_r$ is unaffected as
$\l_r=1-\a$ when $x_0$ is in the low density region of the kink. 

When $x_0$ is in the high density region of the kink as in the case (b), the density is affected
as in the Fig.\ref{fig1}(b).
In this case,  density in the right of $x_0$ is not affected by
detachment and attachment but the density in the left of $x_0$ is
changed by detachment and attachment.
The high density of the particles induce the excess of the detachment
current to the attachment current for $w_a\sim w_d$.
Thus $\l_{\ell}$ and $\l_r$ should be replaced by $\l_{\ell}=\a$ and
$\l_r=\r_-$  when $x_0$ is in the high density region of the kink.

Now, let us determine the value of $\r_-$ and $\r_+$.
Since the exact solution in the steady state of this model is not known,
we adopt the decouple approximation in which the mean currents
$J_{\ell}$ in the low density region of the kink and $J_r$ in the high
density region of the kink are 
given by $J_{\ell}=\l_{\ell}(1-\l_{\ell})$ and
$J_r=\l_r(1-\l_r)$ respectively.
To estimate $\r_+$ and $\r_-$, we use the current conservation at the position
$x_0$. 
For the case when $x_0$ is in the low density region of the kink,
the current from the left of $x_0$ is $\a(1-\a)$ and the current to the
right of $x_0$ is $\r_+(1-\r_+)$.
The detach current from $x_0$ is $w_d\r_+$ and the attach current
at $x_0$ is $w_a(1-\r_+)$.
Thus the equation of the current balance at $x_0$ is given by
\begin{equation}
\label{consm}
 \a(1-\a)+w_a(1-\r_+)=w_d\r_++\r_+(1-\r_+).
\end{equation}
The solution of (\ref{consm}) is simply given by
\begin{equation}
 \r_+=\frac{1+w_a+w_d-\sqrt{(1+w_a+w_d)^2-4w_a-4\a(1-\a)}}{2}.
\end{equation}
Here we use the condition that $\r_+$ should be reduced to (\ref{fden})
for $w_a=w_d=0$.
Similarly, the current balance equation of $\r_-$ is written as
\begin{equation}
\label{consp}
 \r_-(1-\r_-)+w_a(1-\r_-)=w_d\r_-+\a(1-\a),
\end{equation}
and its solution is given by
\begin{equation}
 \r_-=\frac{1-w_a-w_d+\sqrt{(1-w_a-w_d)^2+4w_a-4\a(1-\a)}}{2}.
\end{equation}

From eq.(\ref{vandd}) with the consideration around Fig.1, the drift
velocities $V_{\pm}$ are given by,
\begin{eqnarray}
 V_+=\a-\r_+ \quad {\rm and} \quad
 V_-=1-\a-\r_-.
\end{eqnarray}
Similarly the diffusion constants $D_{\pm}$ are given by
\begin{eqnarray}
 D_+=\frac{\a(1-\a)+\r_+(1-\r_+)}{2(1-\a-\r_+)}  \quad {\rm and} \quad 
 D_-=\frac{\a(1-\a)+\r_-(1-\r_-)}{2(\r_--\a)}.
\end{eqnarray}
Here the quantities with the suffix $+/-$ represent those in the
right/left of $x_0$.
Using $\l_{\ell}=\a,\l_r=\r_-$ for $x<x_0$ and
$\l_{\ell}=\r_+,\l_r=1-\a$ for $x>x_0$, 
the steady solution of the Fokker-Planck equation can be written as, 
\begin{equation}
\label{stsol}
 P(x)=C'\exp\left[\frac{V_-}{D_-}(x-x_0)\th(x_0-x)+\frac{V_+}{D_+}(x-x_0)\th(x-x_0)\right],
\end{equation}
where $C'$ is a normalization constant, and $\th(x)=1$ for $x>0$ and
$\th(x)=0$ otherwise.

For $w=w_a=w_d$, the relations
\begin{equation}
\label{kome}
\r_-=1-\r_+ \quad,\quad |V_+|=|V_-| \quad,\quad D_+=D_- 
\end{equation}
are satisfied.
Thus the solution (\ref{stsol}) becomes
\begin{equation}
\label{mome}
 P(x)=C'e^{-\frac{V}{D}|x-x_0|},
\end{equation}
where $V=|V_-|=|V_+|$ and $D=D_+=D_-$.
From the comparison of eq.(\ref{eqsolb}) with eq.(\ref{mome}), 
the potential energy $U(x)$ is given by 
\begin{equation}
\label{poten}
 U(x)=V|x-x_0|.
\end{equation}
This is one of the main results in this paper.
The validity of our theory will be confirmed by Monte-Carlo simulation
in section 3.

\subsection{Many sites of detachment and attachment}

Here we generalize the model in the previous subsection to a model with
many sites of detachment and attachment.
To ensure the domain wall picture, we assume that $w\doteq w_a=w_d$ are small and the
relations eq.(\ref{kome}) are satisfied.
Substituting the expansion $\r_+$ by $w$ that
\begin{equation}
\label{lin}
 \r_+\simeq \a+w+\cdots,
\end{equation}
we obtain
\begin{equation}
V=w 
\end{equation} 
In this linear regime, we can obtain the probability function of the
kink in many sites of detachment and attachment.

Let us consider the system with $N$ sites of detachment and attachment. 
Thus the system is divided by $N+1$ segments in which each segment is
bounded by the sites of detachment and attachment or the boundaries.

When the kink is located in the $j$th segment, the number of sites of
detachment and attachment in the left of the kink is $j-1$.
For each site of detachment and attachment, the current should be conserved.
The density increases by
$w$ in the low density region as in the eq.(\ref{lin}) in each
segment.
Thus the density $\l_{\ell}$ in the $j$-th segment becomes
$\l_{\ell}=\a+w(j-1)$.
On the other hand, the number of
sites of detachment and attachment in the right of the kink is $N-j+1$.
Thus the density $\l_r$ in the $j$-th segment is $\l_r=1-\a-w(N-j+1)$.
Thus, the drift velocity $V_j$ and the diffusion
constant $D_j$ in the segment $j$ are respectively given by 
\begin{equation}
 V_j=w\left\{N-2(j-1)\right\},
\end{equation}
and 
\begin{equation}
\label{diff}
 D_j =\frac{2\a(1-\a)+(1-2\a)wN+w^2((N-j)^2+j^2)}{2(1-2\a-wN)}.
\end{equation}
Let us omit $j$-dependence in the eq.(\ref{diff})
by neglecting the term proportional to $w^2$.
Thus the diffusion coefficient is reduced to 
\begin{equation}
\label{diff2}
 D=\frac{2\a(1-\a)+(1-2\a)wN}{2(1-2\a-wN)},
\end{equation}
which is independent of $j$.
From eq.(\ref{eqsolb}), the kink probability distribution $P_j(x)$ in
the segment $j$ is given by
\begin{equation}
\label{distri}
 P_j(x)=C_j\exp\left[-\frac{2w(N-2(j-1))(1-2\a-wN)}{2\a(1-\a)+(1-2\a)wN}(x-x_j)\right],
\end{equation}
where $C_j$ is a constant determined from the normalization and 
the compatibility relation 
\begin{equation}
\label{conn}
 P_j(x_j)=P_{j+1}(x_j).
\end{equation}
Thus we obtain the formula applicable to any number of sites of
detachment and attachment.
It should be noted that the denominator in the exponential function in
eq.(\ref{distri}) contain $wN$.
This is because we simply put $D$ in eq.(\ref{diff2}) into eq.(\ref{eqsolb}).
It may be controversial to contain $wN$ in the denominator as the
systematic approximation, but the expression gives us the accurate
result as will be shown later.

This method is also applicable to the PFF model\cite{PFF},
where the number of detachment and attachment sites $N$ is $N=L$, the
segment length is $1$, and the position of the site of detachment and
attachment $x_j$ is $x_j=j$.
Thus the distribution (\ref{distri}) is reduced to $P_j(j)=C_j$. 
We write $P_j=P_j(j)$ for the simplification.
The compatibility relation (\ref{conn}) becomes
\begin{equation}
 C_j=C_{j-1}\exp\left[\frac{w(L-2(j-1))}{D}\right],
\end{equation}
and this recursion relation gives
\begin{equation}
 C_j=C_1\exp\left[\frac{w}{D}\sum_{m=1}^{j-1}(L-2m)\right].
\end{equation}
Thus the distribution is given by
\begin{equation}
\label{dbae}
 P_j=C_1\exp\left[\frac{w}{D}\left\{-(j-\frac{L+1}{2})^2+\frac{(L-1)^2}{4}\right\}\right].
\end{equation}

This result can be also derived by the superposition of the potential.
The harmonic potential is realized by the
superposition of the potential (\ref{poten}) as
\begin{equation}
 U_{PFF}(j)=\sum_{x_0=1}^L w|j-x_0|
\simeq w \left(j-\frac{L+1}{2}\right)^2+\cdots.
\end{equation}
Thus the distribution of the kink position is given by
\begin{equation}
\label{pfff}
 P_{PFF}(j)=C^{''}\exp\left[-\frac{w}{D}(j-\frac{L+1}{2})^2\right],
\end{equation}
where $D$ is given by eq.(\ref{diff2}), and the probability
distribution function of the kink position (\ref{pfff}) is identical to
eq.(\ref{dbae}).

\section{Simulations}

Now let us check the quantitative accuracy of our theoretical argument.
We compare our analysis with the results of Monte-Carlo simulations.
The simulation is carried out by the random update scheme \cite{SP,RSSS},
which is realized by choosing the bond between two
neighbouring site randomly and move the particle in the chosen sites
stochastically.

We use the motion of the second class particle which is a tracer
particle of the kink position to detect the kink
position \cite{FKS,FE,FF}.
If we write $0$ for a hole (empty site), $1$ for a particle and $2$ for
the second class particle, the second class particle moves as
$(2,0)\ri(0,2)$ and $(1,2)\ri(2,1)$.
The hole is moved to the left of the second class particle, and the
particle is moved to the right of the second class particle.
Thus the second class particle is positioned between the low density
region of particles and the high density region of particles.
Thus the second class particle can detect the kink position. 
We compare the probability function of the kink position obtained by
the simulation with the distribution function (\ref{stsol}).

\begin{figure}
\begin{center}
\includegraphics[scale=0.39]{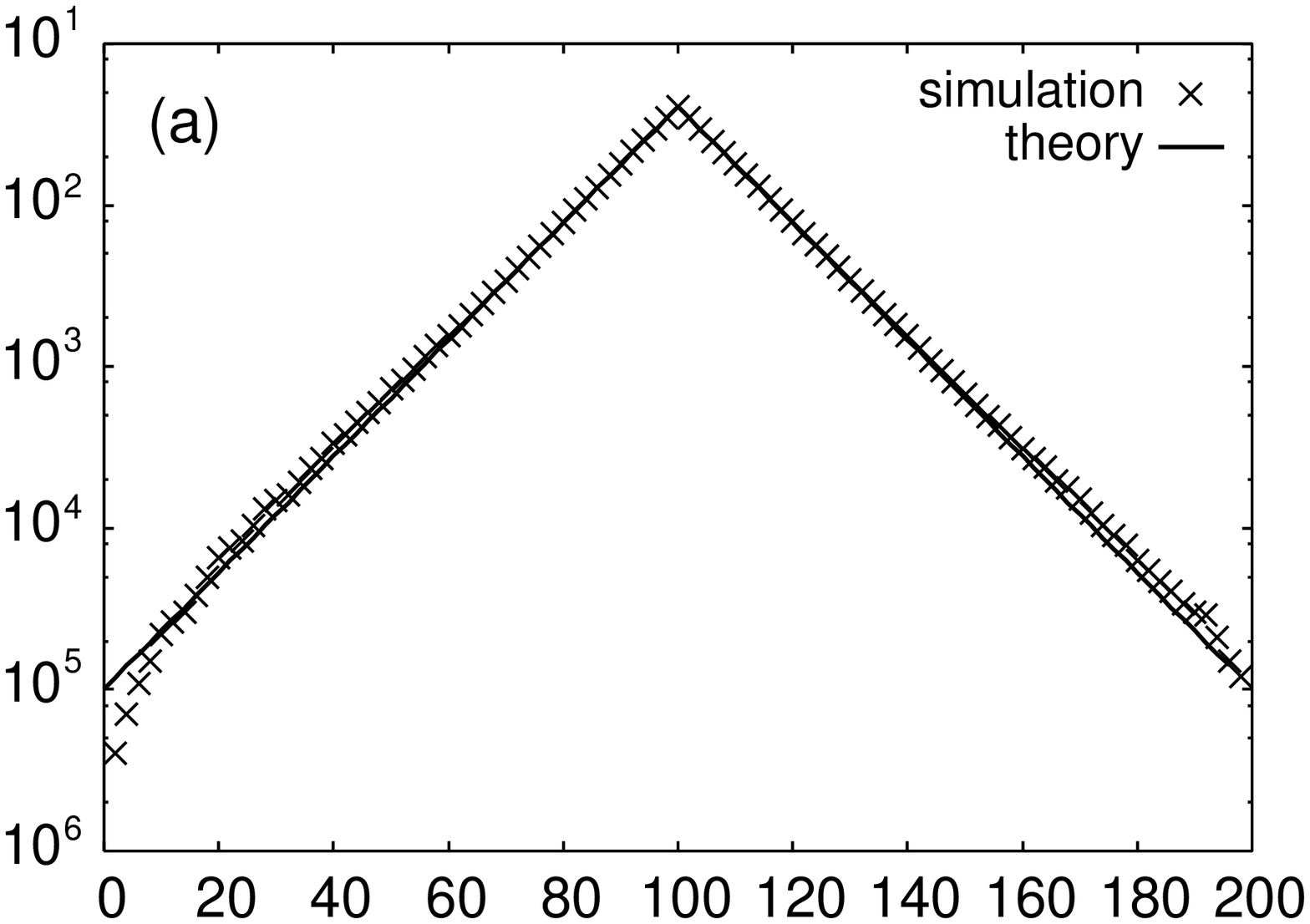}
\includegraphics[scale=0.39]{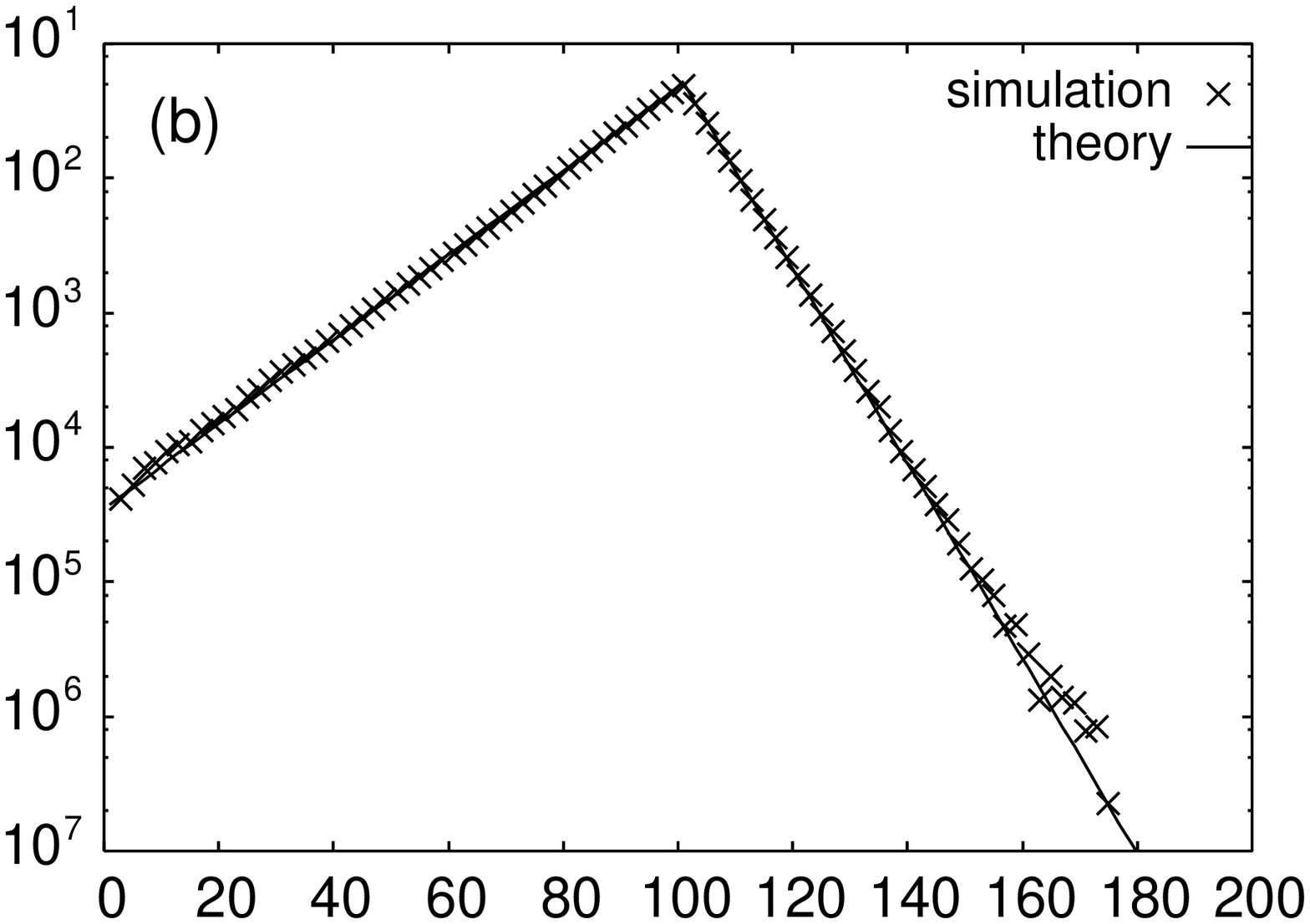}
\end{center}
\begin{flushleft}
\caption{Comparison of the distribution of the kink position between the
 simulation ($\times$) and the solution of the Fokker-Plank
 equation (solid line).The horizontal axis is the kink position and the
 vertical axis is the distribution function of the kink position in the
 steady state plotted in the semi-log scale.
 The parameters used are $\a=\b=0.1$ and $w_a=w_d=0.01$ for (a), and
 $\a=\b=0.1,w_a=0.02$ and $w_d=0.01$ for (b). We set the system length
 $L=200$ and the position of detachment and attachment occur at $x_0=100$.}
\label{fig3}
\end{flushleft}
\end{figure}
As shown in Fig.\ref{fig3}, we obtain good agreement between the
simulation and the theoretical results in eq.(\ref{stsol}).
We plot the results of our simulation by $\times$ and the solution of the Fokker-Plank
 equation by the solid line.
The horizontal axis is the kink position and the
 vertical axis is the probability distribution function of the kink
 position in the steady state.
In Fig.\ref{fig3}(a), the parameter is set to be $w_a=w_d$ and
the distribution function is symmetric around $x=x_0$.
In Fig.\ref{fig3}(b), the parameter is set to be $w_a\neq w_d$
and the distribution function is asymmetric around $x=x_0$.
In both cases, the boundary parameters are set to be $\a=\b=0.1$ and the
system length $L$ is $L=200$.
\begin{figure}
\begin{center}
\includegraphics[scale=0.39]{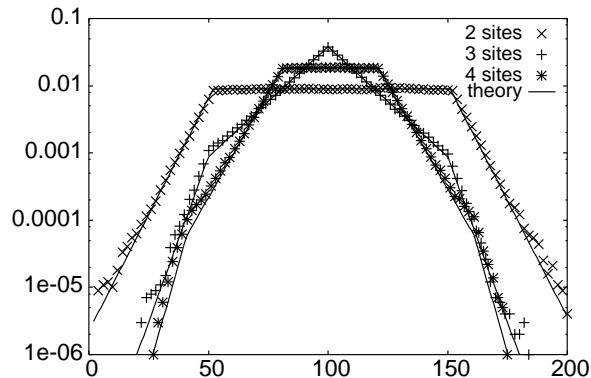}
\end{center}
\begin{flushleft}
\caption{Comparison of the probability distribution of the kink position between the
 simulation and the solution of the Fokker-Planck
 equation (solid lines) in eq.(\ref{distri}) for the
 case which has $2(\times)$, $3$(+) and $4(*)$ sites of detachment
 and attachment.
The horizontal axis is the kink position and the
 vertical axis is the probability function of the kink position in the
 steady state plotted in the semi-log scale.
The parameters used are
 $\a=\b=0.1$ and $w=0.01$, and the positions of detachment and attachment
 are $x_1=50,x_2=150$ for $N=2$,
 $x_i=50i,(i=1,2,3)$ for $N=3$ and $x_i=40i,(i=1,2,3,4)$.
We set the system length $L=200$. }
\label{fig4}
\end{flushleft}
\end{figure}
In Fig.\ref{fig4}, we compare the theoretical results with
the results of our simulation in cases of $2$, $3$ and $4$ sites of
detachment and attachment.
The parameters used in this case are
 $\a=\b=0.1$ and $w=0.01$, the position of detachment and attachment are
 $x_1=50,x_2=150$ for $N=2$,
 $x_i=50 i,(i=1,2,3)$ for $N=3$ and $x_i=40 i,(i=1,2,3,4)$.
The results of our simulation are plotted by $\times$ for $N=2$, $+$ for $N=3$
and $*$ for $N=4$, and the theoretical results are plotted in the solid
lines.
\begin{figure}
\begin{center}
\includegraphics[scale=0.39]{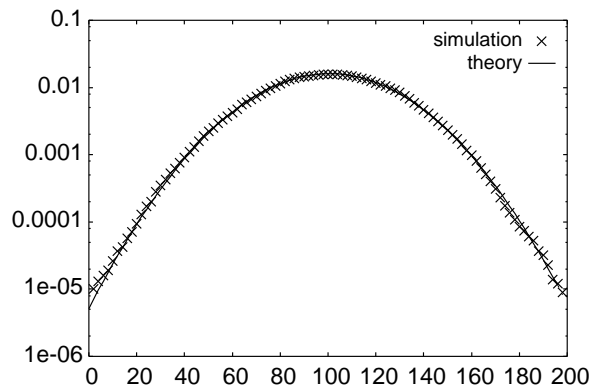}
\end{center}
\begin{flushleft}
\caption{Comparison of the probability distribution of the kink position between the
 simulation ($\times$) and the solution of (eq.(\ref{distri}))the Fokker-Planck
 equation (solid line) of the PFF model.
 The horizontal axis is the kink position and the
 vertical axis is the distribution function of the kink position in the
 steady state plotted in the semi-log scale. 
The parameters used are $\a=\b=0.1,w=0.0001$ and $L=200$. }
\label{fig5}
\end{flushleft}
\end{figure}
In Fig.\ref{fig5}, we compare the theoretical results with
the simulation results in the PFF model.
The parameters at the boundaries are $\a=\b=0.1$ and $w=0.0001$.
The results of our simulation give quite well agreement of the theoretical
prediction in eq.(\ref{distri}) in all cases.

\section{Conclusion}

We have demonstrated that the kink motion in TASEP with detachment and
 attachment can be described by the Brownian motion under the influence
 of the attractive force from detachment and attachment sites.
We have obtained the attractive potential to the kink and the diffusion
 constant of the kink.
We demonstrate that the superposition of the potentials of our model
 gives good results 
 for any number of sites of detachment and attachment when the
 rates of detachment and attachment are small.
We compare our result with the simulation and have confirmed that our
 theoretical prediction of our theory gives quantitatively correct results.
We also explain the reason why the kink in PFF model feels a harmonic
 potential \cite{EJS,JS}, and succeed the quantitative estimation of the potential.

We would like to thank S.Takesue for fruitful discussion.
This work is partially supported by the Grant-in-Aid for Ministry of
Education, Science and
Technology(MEXT), Japan (Grant No. 18540371), the Grant-in-Aid for the 21st century COE
'Center for Diversity and Universality in Physics' from MEXT, Japan, and
the Grant-in-Aid of Japan Space Forum.

\end{document}